\documentclass{article}
\usepackage{PRIMEarxiv}
\usepackage[utf8]{inputenc} % allow utf-8 input
\usepackage[T1]{fontenc}    % use 8-bit T1 fonts
\usepackage{hyperref}       % hyperlinks
\usepackage{url}            % simple URL typesetting
\usepackage{booktabs}       % professional-quality tables
\usepackage{amsfonts}       % blackboard math symbols
\usepackage{bm}
\usepackage{nicefrac}       % compact symbols for 1/2, etc.
\usepackage{microtype}      % microtypography
\usepackage{lipsum}
\usepackage[running]{lineno}
\usepackage{multicol}

\usepackage{color}
\definecolor{darkgreen}{rgb}{0.0,0.4,0.0}

\usepackage{fancyhdr}       % header
\usepackage{graphicx}       % graphics

\title{
Comparison of two 
artificial neural networks trained for the surrogate modeling of stress 
in materially heterogeneous elastoplastic solids
}

\author{
  Sarthak Kapoor \\
  RWTH Aachen\\
  \texttt{sarthak.kapoor@rwth-aachen.de} \\
  \And
  Jaber Rezaei Mianroodi \\
  MPIE Düsseldorf\\
  \texttt{j.mianroodi@mpie.de} \\
  \And
  Mohammad Khorrami \\
  MPIE Düsseldorf\\
  \texttt{m.khorrami@mpie.de} \\
  \And
  Nima S. Siboni \\
  MPIE Düsseldorf\\
  \texttt{nima.siboni@gmail.com} \\
  \And
  Bob Svendsen \\
  MPIE Düsseldorf, RWTH Aachen\\
  \texttt{b.svendsen@mpie.de} \\
  \And
}

\begin{document}
\maketitle
% \linenumbers

\begin{abstract}

%In this work, we systematically compared convolutional-neural-network-based U-Net and Fourier neural operators (FNO) as surrogate architectures for simulations in micromechanics. Both these approaches were adapted to learn mappings from material properties fields to stress fields, given by the first Piola-Kirchhoff stress tensor, under uniaxial tensile loading. We found that both networks give tremendous speedups of almost 1000 and 2500 for U-Net and FNO respectively, as compared to the spectral solver used for generating the data. Further, we found that FNO gives much lower normalized mean absolute errors of about 0.25-0.40\%, as compared to U-Net which was found to be 1.41-2.15\% for different stress components when tested on similar geometries as the training data. We also investigated how well the two approaches generalize when tested with different resolutions and geometries and found that error in  FNO only scaled with the percentage of grain boundary region, while the error in U-Net always increased as the test data moved further away from training data. We further tested the networks with different aspect ratios of the boundary box and found the calculations from both networks to be plagued with artifacts in the case of aspect ratios smaller than one. Finally, we put forth intuitive reasoning for the different error behavior of the two networks and recommend using FNO for multi-output spatial field regression. 

%\color{blue} 

The purpose of this work is the systematic comparison of the application of two artificial neural networks (ANNs) to the surrogate modeling of the stress field in materially heterogeneous periodic polycrystalline microstructures. The first ANN is a UNet-based convolutional neural network (CNN) for periodic data, and the second is based on Fourier neural operators (FNO). Both of these were trained, validated, and tested with results from the numerical solution of the boundary-value problem (BVP) for quasi-static mechanical equilibrium in periodic grain microstructures with square domains. More specifically, these ANNs were trained to correlate the spatial distribution of material properties with the equilibrium stress field under uniaxial tensile loading. The resulting trained ANNs (tANNs) calculate the stress field for a given microstructure on the order of 1000 (UNet) to 2500 (FNO) times faster than the numerical solution of the corresponding BVP. 

For microstructures in the test dataset, the FNO-based tANN, or simply FNO, is more accurate than its UNet-based counterpart; the normalized mean absolute error of different stress components for the former is 0.25-0.40\% as compared to 1.41-2.15\% for the latter. Errors in FNO are restricted to grain boundary regions, whereas the error in U-Net also comes from within the grain. In comparison to U-Net, errors in FNO are more robust to large variations in spatial resolution as well as small variations in grain density. On other hand, errors in U-Net are robust to variations in boundary box aspect ratio, whereas errors in FNO increase as the domain becomes rectangular. Both tANNs are however unable to reproduce strong stress gradients, especially around regions of stress concentration.
% \color{darkgreen} 
% Following the now standard convention, 
% I assume the total data set is split into training (80\%) and test 
% (20\%) subsets? -- we used 800:100:100 split for train:validation:test
% \color{black}
% \color{magenta} 
% tANN modeling of microstructures with morphologies and grain sizes not in the training data show that the error of the FNO-based 
% tANN is restricted to the grain boundary regions, whereas the error in 
% U-Net always increased as the test data moved further away from 
% training data. 
% \color{darkgreen} 
% By "not in the training data", do you mean in the test data set? 
% How do the training set and test errors compare for the U-Net- and FNO-based tANNs? 
% \color{black}
% In addition, tANN modeling of non-periodic input suffers from increased error.
% \color{darkgreen} 
% Since your training and test data sets contain only periodic data, 
% non-periodic data lie "outside" of your total data set, and this is 
% hardly surprising. 
% \color{black}

\end{abstract}


\section{Introduction}

%\color{blue}

The mechanical response of materials depends heavily on the microstructure and 
its morphology. In the case of polycrystalline metallic materials, for example, 
this is based on grains of different sizes as well as defects like inclusions 
and precipitates 
\cite{Hall1954,Gladman2013_ppt_hardening,Suresh2009_nanoscale_boundaries}. 
%However, the process of experimentally pinpointing the right microstructures 
%quickly runs into a combinatorial explosion of possible variations, making 
%it both cumbersome and resource-draining. 
Computational modeling of such microstructure, its evolution, and the corresponding material behavior is often based on phase field, finite element (FE), 
or spectral, methods, which complement and even reduce experimental efforts  
\cite{MOELANS2008268,KOCHMANN201689,SHANTHRAJ2016167,SUN20013007,
EISENLOHR201337,SHANTHRAJ201531}. 
Being based on the numerical solution of initial boundary-value problems (IBVPs), 
such modeling generally suffers from high computational costs, especially for 
increasing model non-linearity, coupled equation system size, or physical 
system size. 

Because of this, alternative approaches based on artificial neural networks (ANNs) 
and machine learning (ML) have attracted much attention in the field of material 
mechanics simulation, in particular, due to the possibility of a tremendous reduction 
in computational cost. In these approaches, one trains an ANN using input and 
results from the numerical solution of physical IBVPs. The resulting trained ANN 
(tANN) is then employed for surrogate computational modeling of input orders of 
magnitude faster than the numerical solution of the corresponding IBVP. 

%\color{black}

Convolutional Neural Networks (CNN) have emerged 
as a promising choice due to their ability to learn features in 
structured grid-like data settings. Yang et al.~\cite{YANG2019335} 
showed the superiority of the CNN-based deep learning approach over 
other ML approaches based on feature extraction to calculate the 
elastic strain field of three-dimensional high-contrast elastic composites. 
Yang et al.~\cite{YANG2021104506} employed a conditional generative 
adversarial neural network (cGAN) to calculate stress and 
strain tensor fields in two-phase heterogeneous elastic materials. 
Mianroodi et al.~\cite{Mianroodi2021} used a U-Net-based CNN to 
calculate the von Mises stress field in highly heterogeneous grain 
microstructures. 

One limitation of NNs based on neurons or perceptrons 
is the fact that they approximate (only) functions between 
finite-dimensional spaces \cite{HORNIK1989359,ZHOU2020787}. 
As shown also in the current work, 
one consequence of this is a dependence of the tCNN on the spatial 
resolution of the training data. 
In contrast, NNs based on neural operators approximate functionals 
between function spaces \cite{LuLu2020_DeepONet, Li2020_NeuralOperator}, 
resulting in training independent of the spatial resolution of the data. 
This ability has been referred to as "single-shot superresolution" 
in the literature as one can train the network at a lower resolution 
and test it at a higher resolution without re-training. 
Li et al.~\cite{Zongyi2020-FNO} developed a such neural operator 
based on Fourier methods. The resulting Fourier neural operator (FNO) 
was applied by them to approximate the solution of partial differential 
equations from fluid dynamics and showed that indeed that the training 
and test errors are independent of the spatial resolution of the data. 
More recently, FNO-based NNs and related DeepONet-based ones 
\cite{LuLu2020_DeepONet} have been applied to problems in micromechanics. 
Rashid et al.~\cite{Rashid2022_composites_usingFNO} used the FNO 
framework to calculate tensorial stress-strain fields in two-component 
2D digital structures. 
Oommen et al.~\cite{Oommen2022_twoPhaseMicrostructure_neuralOps} employed 
DeepONet in an autoencoder architecture to calculate microstructural 
evolution based on data from phase-field simulations. 

The purpose of the current work is a detailed comparison of 
a UNet-based CNN and FNO-based ANN trained with data from 
finite-deformation micromechanical modeling of synthetic 2D grain 
microstructures. Both networks are trained to calculate the 
9 Cartesian components \(P_{11},P_{12},\ldots,P_{32},P_{33}\)
of the first Piola-Kirchhoff stress tensor \(\bm{P}\) for 
isotropic, perfect elasto-plastic grains. Synthetic grain microstructures 
are generated with the help of Voronoi tessellation. 
The goal of the study is to compare the performance of the two 
approaches in the terms of accuracy (both local or average behavior 
over the field), interface and boundary resolution, as well as 
the relative increase in numerical efficiency. 
Additionally, we investigate the extent to which these approaches 
generalize for variable spatial resolution, boundary box aspect ratio, 
microstructure morphology. 

The current work begins in Section \ref{sec:ArcNet} with a summary 
of the network architectures for the UNet- and FNO-based NNs. This 
is followed by a discussion on data generation in Section \ref{sec:dataset} 
and network training and testing in Section \ref{sec:TestTra}. In Section \ref{sec:Res} we discuss the resulting inferences from the two networks and finally close with a conclusion in Section \ref{sec:ConSum}.
% \color{darkgreen}
% As I have started to do, you need to 
% summarize the structure of the paper at the end of the
% introduction. 

\color{black}

\section{Network Architecture} 
\label{sec:ArcNet}

\subsection{U-Net}
U-Net is a popular convolutional neural network and was originally developed for application in biomedical imaging \cite{ronneberger2015unet}. It encodes the input \(n\)-times into lower spatial dimensions and then decodes it back \(n\)-times to recover the original spatial dimensions. Throughout the network, multiple convolution kernels learn representations at different spatial dimensions/resolutions. 

In our implementation of U-Net using the TensorFlow library 
\cite{tensorflow2015-whitepaper}, 
the input has three channels, one for each material property, 
and the output has five channels, one for each stress component 
(see Section \ref{sec:dataset} to know why we consider five stress 
components). As a fully-convolutional implementation, the network can handle input with any spatial dimension. For the encoding steps, we use a max-pooling layer with \((2,2)\) size followed by a batch-normalization layer, whereas for the decoding step, we use upsampling layer that performs bilinear interpolation with \((2,2)\) size. This way, we need an equal number of encoding and decoding steps, which were chosen to be four. The number of channels of the activation maps increases during the encoding steps and decreases during the decoding step. This corresponds to learning more/fewer convolution filters at lower/higher spatial dimensions. 

\begin{figure}[ht]
    \centering
    \includegraphics[height=8cm]{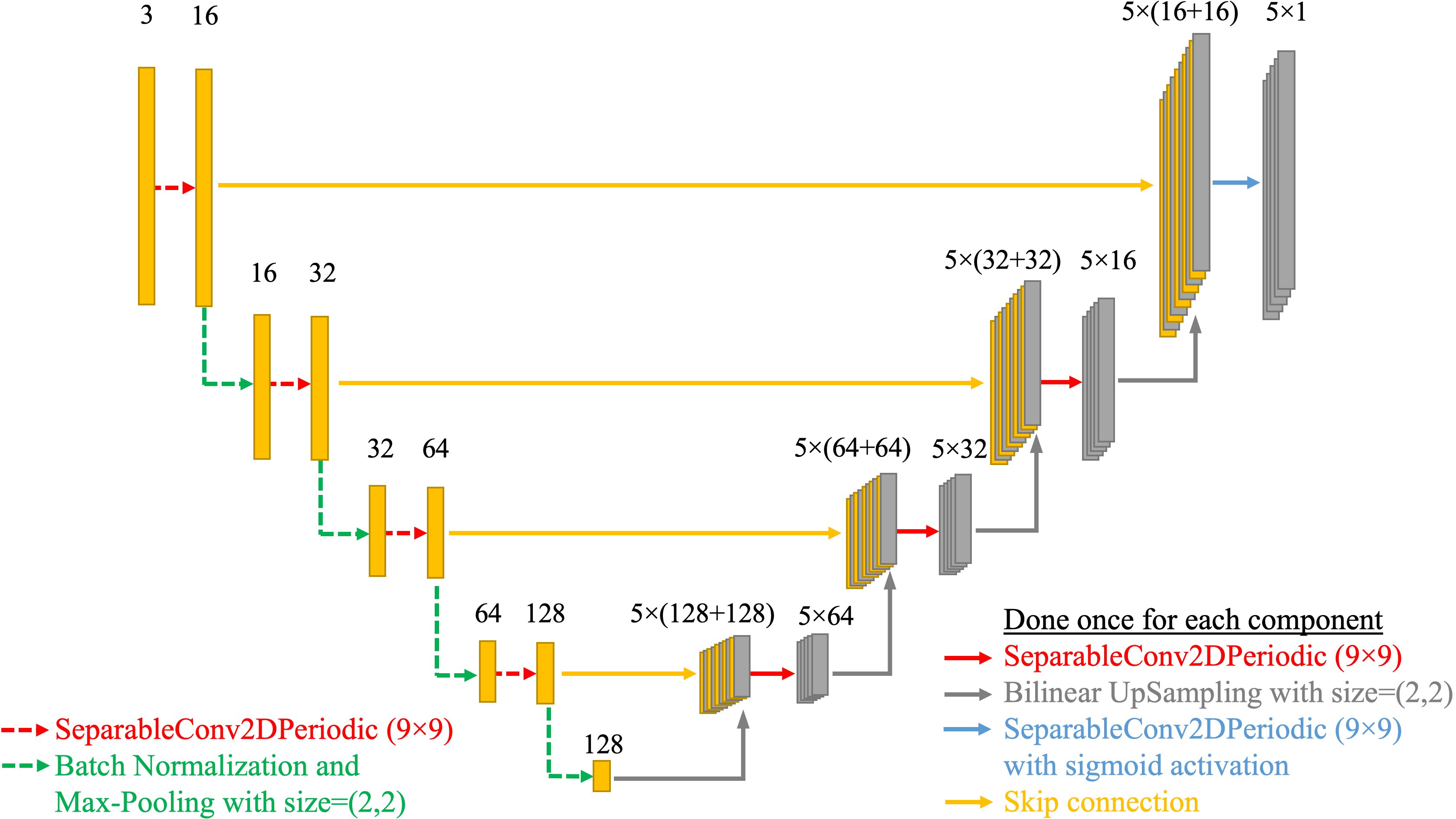}
    \caption{Network architecture of U-Net where the division of network happens from the first decoding stage to generate multiple output fields. Each activation map is labeled with its number of channels; spatial dimensions get halved after each encoding stage and doubled after each decoding stage.}
    \label{fig:U-Net architecture}
\end{figure}

We modified the convolution filter to consider periodic boundary conditions. When it processes the boundaries of the activation maps, it considers sufficient periodic padding to generate outgoing activation maps with identical spatial dimensions. These "separable periodic convolutions" learn \(9\times9\) filters at each encoding/decoding step. Finally, our architecture utilizes skip connections \textemdash after every decoding step, the network concatenates outgoing activation maps with corresponding activation maps from encoding stages having identical spatial dimensions. This preserves the higher resolution information that might be lost during encoding (see Figure \ref{fig:U-Net architecture}).

We found that not using nonlinear activation functions after convolution layers gave better accuracy for multi-output U-Net. However, we used sigmoid activation for the output layer to have the stress calculation values in \([0,1]\), which is also the normalized stress range we use during training and testing. To generate multiple output fields, the network branches into as many subnetworks as the number of output fields required. We worked with three such adaptations where the network either branches from the first encoding layer, the first decoding layer, or the output layer. The corresponding results from these adaptations have been discussed in Section \ref{subsec:MultAdapt}.

\subsection{Fourier neural operator}
FNO implements operator-based learning by learning linear transforms in Fourier space. Parameterization in Fourier space allows the network to learn the underlying frequency information in the given data distribution. To much extent, this is independent of the spatial resolution of the \(2\)D data, provided that features to be learned are frequent enough in the given resolution, and corresponds to learning a family of functions, rather than a single function specific to training resolution. Further, Fourier layers stack onto each other via non-linear activation functions, which allows them to learn highly non-linear operators. Periodicity is inherent to Fourier space, therefore this kind of parameterization works well for periodic boundary conditions. Additionally, by adding a parallel real-space transform (bias) in the Fourier layer, non-periodic information is also picked by these layers. 

\begin{figure}[ht]
    \centering
    \includegraphics[height=7.5cm]{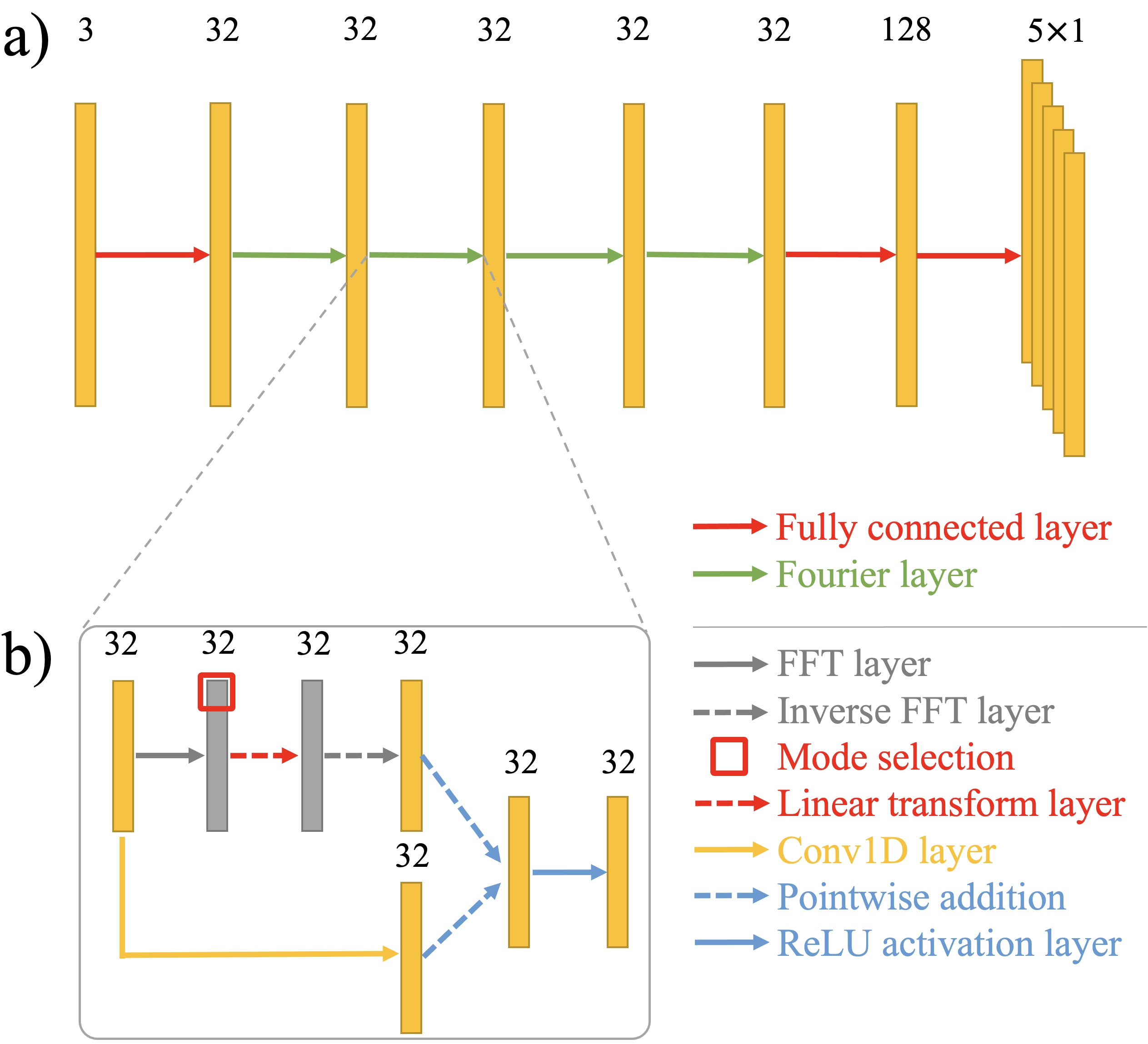}
    \includegraphics[height=7.5cm]{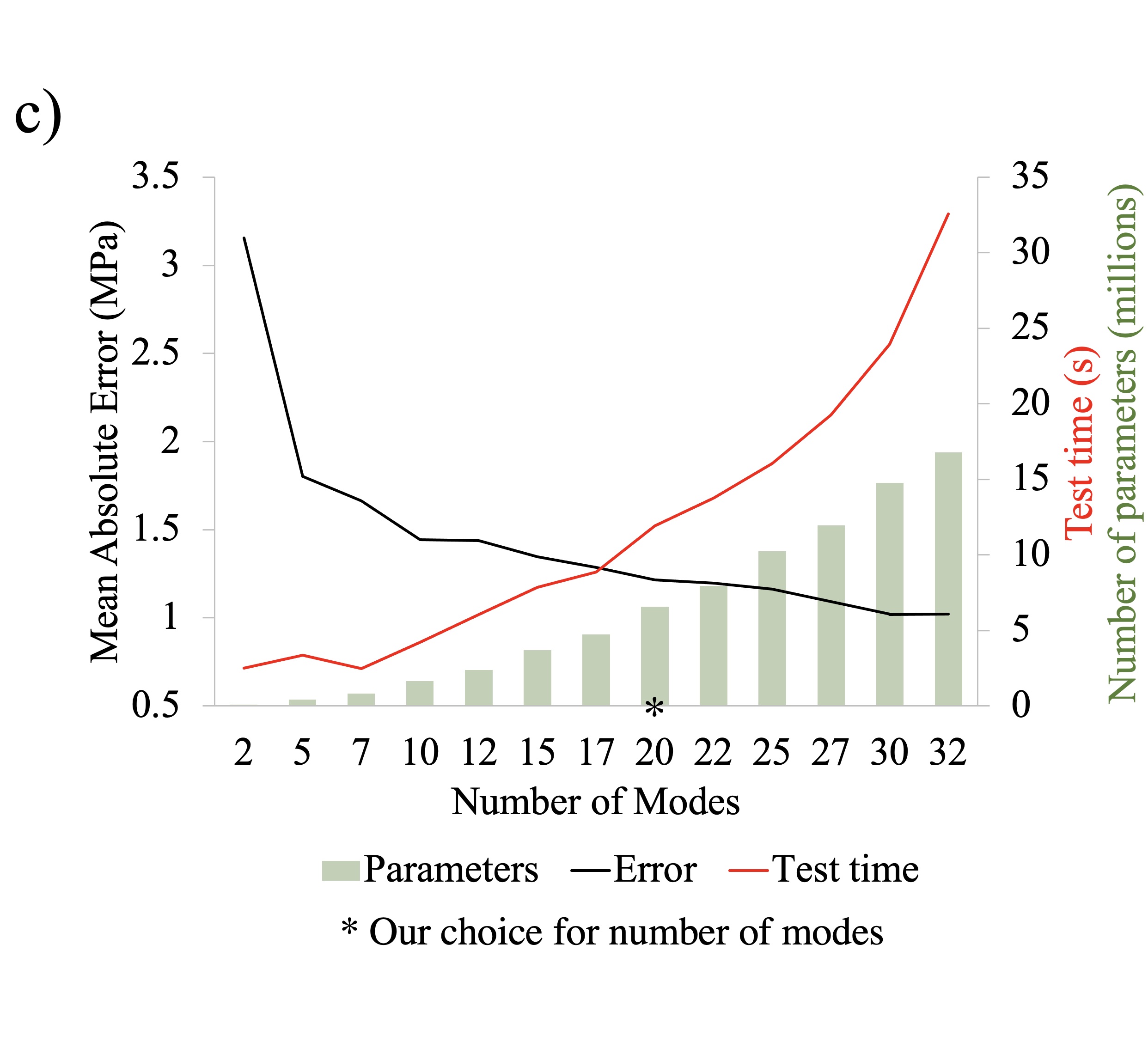}
    \caption{\textbf{a)} Network architecture of FNO where the division of network for multiple output fields happens from the output layer. \textbf{b)} Internal working of FNO layer \textemdash here the dimensions of activation maps remain the same after every layer; linear transform is done for a selected number of Fourier modes and values of other modes are set to zero before inverse FFT. \textbf{c)} Test error, test time, and number of trainable parameters for single-output FNO with varying number of modes (for stress component \(P_{11}\) at \(64\times64\) resolution)}
    \label{fig:FNO architecture}
\end{figure}

To implement operator learning, we referred Fourier neural operator (FNO) network as devised by Li et. al, \cite{Zongyi2020-FNO} and extended it to generate multiple output channels. The network has been implemented using PyTorch (version 1.6) framework \cite{PyTorch}, and just as for U-Net, the network takes three-channel input and gives five-channel output. It uses fully connected layers to raise/decrease the number of channels of activation maps. We used a fixed number of \(32\) channels for every Fourier layer (four layers stacked one after the other), with ReLU activation between each of them. As the network does not alter the spatial dimensions throughout its length, it can handle input with any spatial dimensions. We tested three multi-output adaptations where the branching of the network happens at either the first Fourier layer, the third Fourier layer, or the output layer (see Figure \ref{fig:FNO architecture}a-b). 

To transform back and forth between real and Fourier space, we use fast Fourier transforms. However, not all the Fourier modes are used in the linear transform within the Fourier layer. First \(n\)-modes in each spatial dimension are preserved, while others are set to zero. This mode selection is an important hyperparameter for FNO, which at large, decides the size of the parameter space. We trained and tested FNO for different numbers of preserved modes using a low-resolution training set (\(64\times64\)), and found the test error to reduce as the number of preserved modes increased. But this is accompanied by an almost quadratic increase in parameter space. Nevertheless, this provides an easy-to-tune hyperparameter to get more accuracy at cost of computation time. We used \(20\) as the number of preserved modes for all the experiments (see Figure \ref{fig:FNO architecture}c).

Interestingly, for the same dataset, we would need more than \(20\) modes to faithfully preserve the information in a standard forward-and-inverse Fourier transform (no learning involved). This indicates that even with fewer modes, the network is able to faithfully reproduce the stress fields by the virtue of learning important features/statistics from the training dataset.  

\section{Data generation}
\label{sec:dataset}

The data for training and testing of the above ANNs is obtained from 
micromechanical modeling of periodic 2D synthetic grain microstructures 
assuming each grain is an isotropic, elastic, perfectly plastic material. 
The corresponding periodic boundary value problems have been solved 
numerically with the help of spectral methods implemented in the 
D\"usseldorf Advanced Material Simulation Kit (DAMASK) 
\cite{Roters2019-DAMASK}. 

Synthetic grain microstructures have been generated in a simulation box of 
dimensions \((n_x,n_y,n_z)\)\footnote{One of the spatial dimension only had two discretized points for the purpose of running \(2.5\)D simulation, however, only \(2\)D data was extracted for training and testing of networks.} by randomly setting as many seeds as the required number of grains. We then performed Voronoi tessellation with seeds acting as the centers of the individual regions. By spreading the grains almost uniformly, we were able to mimic equiaxial grain structure, which is common to many real materials. For each grain, material properties (three in total \textemdash Young's modulus, Poisson's ratio, yield stress) were specified. These material properties are homogenous within each grain/phase and are assigned at random from a defined set of values, i.e., \(\{60,80,100,120\}\) for Young's modulus (GPa), \(\{0.1,0.2,0.3,0.4\}\) for Poisson's ratio, and \(\{50,100,150,200\}\) for yield stress (MPa). For all the simulations, the microstructure was subjected to small uniaxial tensile loading in the direction which would translate to vertical \(\updownarrow\) direction with respect to this text. 

We ran the simulations for 20 increments to reach an almost stationary mechanical state under the applied load and extracted the first Piola-Kirchhoff stress tensor \(P_{ij}\) at the final time step. \(P_{ij}\) specifies \((3,3)\) stress tensor at every point, making it a \(4\)-tensor \((3,3,n_y,n_z)\). However, due to uniaxial tensile loading, only five components (\(P_{11}\), \(P_{22}\), \(P_{23}\), \(P_{32}\), \(P_{33}\)) have non-trivial stress distribution; remaining components have homogenous response throughout the microstructure. Therefore, we use the ANNs to calculate only the non-trivial stress components, with the assumption that it could be extended to calculating all stress components when loading is more complex. As the goal of this study was to compare two ML-based approaches for correlative modeling, we could safely work with this simpler situation to establish important comparisons between them.

After postprocessing, the dataset included \(2\)D spatial fields which inherently specify the morphology of the microstructures. The data for the input side of the ANN comprised three spatial fields, each specifying the distribution of one material property at every discretized point. The data for the output side of the ANN comprised five spatial fields, each specifying one non-trivial stress component (see Figure \ref{fig:calculation 256}a-b). For primary training purposes, we generated \(1000\) cases with \(20\)-grain microstructure and \(256\times256\) resolution, whereas to do secondary training for the purpose of hyperparameter selection, we generated a similar dataset at \(64\times64\) resolution. Further, we generated cases with different grain sizes and morphologies, aspect ratios, and resolutions for the purpose of testing and comparison.

\section{Training and testing}
\label{sec:TestTra}

For both ANNs, we randomly split the dataset into \(800\) training cases, \(100\) validation cases, and \(100\) unseen test cases. We used mean absolute error (MAE) for U-Net as the loss function to be minimized. The choice is based on findings by Qi et. al \cite{MAE_for_DNN}, who showed that MAE loss is more robust to noise than mean squared loss (MSE) for training deep neural networks. The weights were initialized using uniform Glorot initialization \cite{pmlr-v9-glorot10a}, which is set by default in TensorFlow \cite{tensorflow2015-whitepaper}. On the other hand, we used MSE as the loss function for FNO, and each layer was initialized with random weights drawn uniformly from \([0,1)\) and scaled by \(\frac{1}{(N_{in} \times N_{out})}\), where \(N_{in}\) and \(N_{out}\) are the number of input and output channels for the given layer, as used by Li et. al \cite{Zongyi2020-FNO}. For both ANNs, we used Adam optimizer \cite{Adam} with learning rates of \(0.001\). By end of the training, the validation losses for both ANNs were fairly saturated.

\paragraph{Normalized mean absolute error}

To compare the performance of ANNs in the calculation of the stress response on average, we aggregated pixel errors from the entire field by averaging them. We extensively used MAE as an evaluation metric while tuning the network architectures for single stress-component calculation. On other hand, we found normalized mean absolute errors (NMAE) to be a better-suited metric than mean absolute or relative errors when working with multiple stress component calculations. As for a given test case, the range of values for each stress component is different, reporting absolute errors (like MAE or MSE) would be misleading when comparing the average errors for different components. In this case, accumulating relative errors at each pixel would be a preferred choice as it would factor out magnitudes of different stress component values. However, as the stress values range from negative to positive values, some of these are close to zero. These regions with almost zero stress values would give huge contributions if we compute relative errors, irrespective of the local error incurred here. Interpretations from average relative errors (like relative MAE) would therefore be misleading. NMAE avoids these issues because, unlike absolute errors, these values are normalized with the range of stress values, which makes NMAE from different stress components comparable. Moreover, this type of normalization does not explode the error values for regions with near-zero stress values. NMAE is given by 
\begin{equation}
    NMAE=  \frac{\frac{1}{mn} \sum^{mn}_i|X_i - Y_i|}{max(X)-min(X)},
\end{equation}
where \(X, Y\) are stress fields from spectral solver and ANN calculation respectively, each being an \(m\times n\) array. Here, the term in the numerator is nothing but MAE. If not mentioned otherwise, average field errors were averaged over \(10\) (or more) test cases for stable estimates. 

\section{Results}
\label{sec:Res}

\subsection{Multi-output adaptation of networks}
\label{subsec:MultAdapt}

\begin{figure}[ht]
    \centering
    \includegraphics[height=5cm]{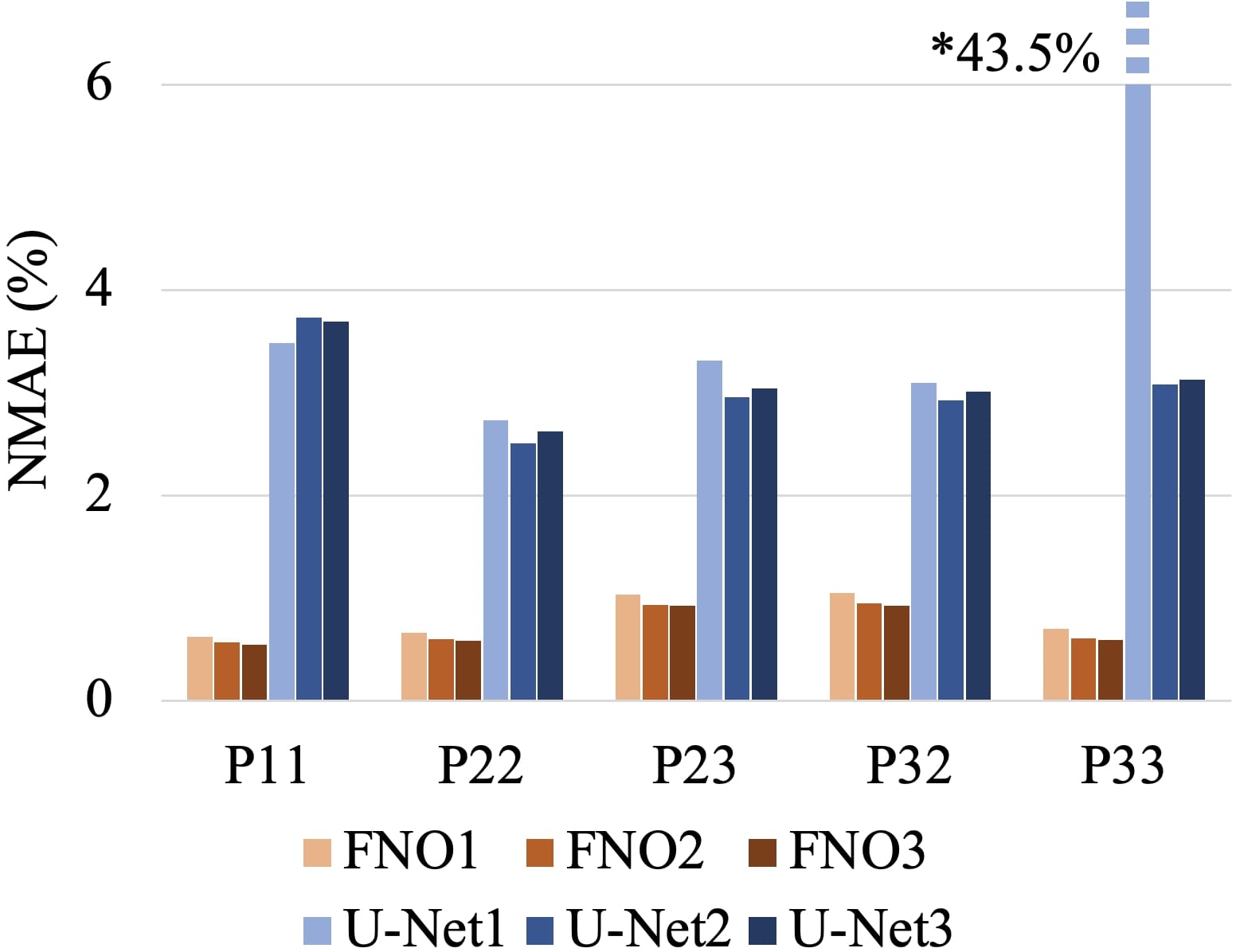}
    \caption{Normalized mean absolute error (taken over 100 cases for each stress component) for multi-output variants of U-Net and FNO. The errors were computed for 5-stress-component calculation at \(64\times64\) resolution.}
    \label{fig:Multiple output variants}
\end{figure}

We adapted both U-Net and FNO to generate multiple output fields by branching them into sub-networks at different stages where each sub-network calculated one stress component, as described in Section \ref{sec:ArcNet}. To compare the calculation errors from different multi-output variants, we trained and tested every network variation at \(64\times64\) spatial resolution of the input and output fields. The selected variants were then re-trained at a finer resolution of \(256\times256\) for further tests. As expected, the number of trainable parameters increases if the branching happens earlier in the network, and so does the network evaluation time (see Table \ref{tab:Multi-output variants}). However, we also observe that increasing the number of parameters does not always improve calculations. As shown in Figure \ref{fig:Multiple output variants}, the error for component \(P_{33}\) using U-Net1, where the branching happens at the first encoding stage, hence having the maximum number of parameters, is an order of magnitude higher than other adaptations.  It might be because of the limitations of training data to find better optima in what is now a bigger parameter space. Nevertheless, prediction errors were also not the least for U-Net with the most delayed branching (U-Net3) due to insufficient parameter space for required learning. A trade-off between these two factors happens when U-Net is branched from the first decoding layer (U-Net2), which led to the lowest errors in four out of five components. In the case of FNO, the trend is more regular where the prediction error increased monotonically for all stress components when the branching of the network happens in earlier layers. FNO3, where the network branching happens from the output layer, gave the lowest errors with the least number of parameters among all variants. Therefore, the last fully connected layer with just \(645\) parameters was sufficient to distinguish between all five components with lower errors than U-Net. We selected U-Net2 and FNO3 variants for further study, and refer to them as U-Net and FNO respectively. 

We already see an improvement in errors going from U-Net to FNO for training at \(64\times64\) resolution, but more interestingly, we also observe that while the number of trainable parameters in FNO is almost \(24\) times higher than U-Net, its inference/test time is still lower. This is because in FNO the learned weights of linear transforms are computed over once per forward pass through the network. On other hand, in U-Net the learned convolution kernels slide over the spatial fields and are computed over multiple times, once after each stride. Therefore, even for a lower number of parameters, the amount of computation required for a forward pass in U-Net is higher.

% \color{red}
% I am unable to find a reliable way of computing FLOP for the two architectures. Issues- TensorFlow profiler probably doesn't account for FLOP in the upsampling layer, and maybe gives erroneous results for custom layers, like our separable periodic convolution. For FNO, I used the profiler from 'thop' library and have a number (MAC=365.691M for 64x64 resolution), but not sure if it's correct.  Any suggestions here will be helpful. Otherwise, we can choose not to report this remark altogether.
% \color{black}

\begin{table}[ht]
 \caption{Test time (averaged over \(10\) cases with 5-component calculation at \(64\times64\) resolution) and the number of parameters for multi-output variants of U-Net and FNO. The rows with bold font represent the U-Net and FNO variants selected for further study.}
  \centering
    \begin{tabular}{l|l|c|c}
    \toprule
    %\multicolumn{2}{c}{Part}                   \\
    %\cmidrule(r){1-2}
    & Variant & Test time (ms) & Number of trainable parameters \\
    \midrule
    U-Net1 & From first encoding stage  & 171 & 329, 250 \\
    \textbf{U-Net2} & \textbf{From first decoding stage}  & \textbf{123} & \textbf{245, 910} \\
    U-Net3 & From output stage  & 73 & 72, 086 \\
    \midrule
    FNO1 & From first Fourier layer  & 677 & 32, 811, 013\\
    FNO2 & From third Fourier layer  & 428 & 19, 695, 365 \\
    \textbf{FNO3} & \textbf{From output layer}  & \textbf{91} & \textbf{6, 562, 821} \\
    \bottomrule
  \end{tabular}
  \label{tab:Multi-output variants}
\end{table}

\subsection{Stress field calculations}\label{subsec:stress field calculations}

\begin{figure}[ht]
    \centering
    \includegraphics[height=7cm]{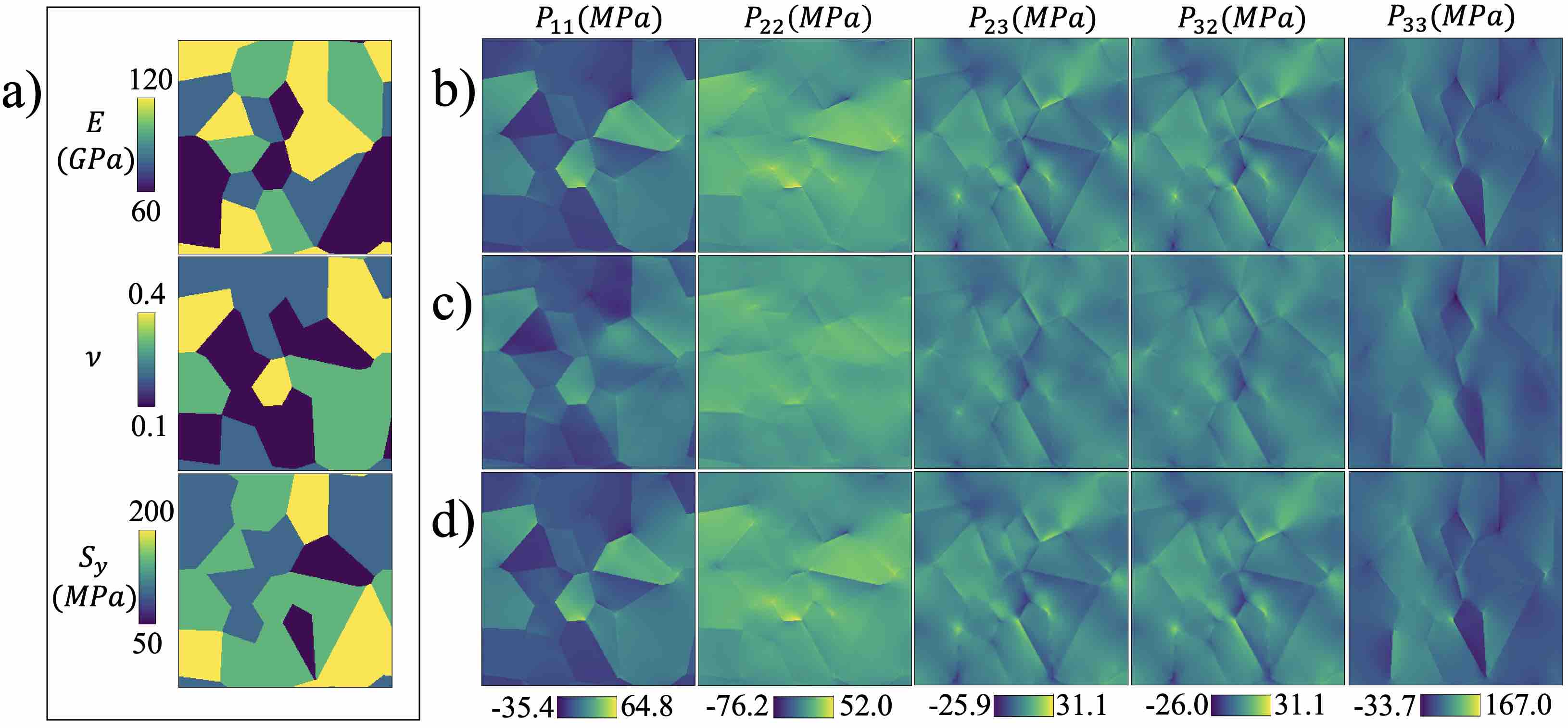}
    \caption{Network calculations on a test case with \(256\times256\) resolution and \(20\) grains (trained at similar geometries) \textemdash \textbf{a)} Material properties as input fields to the network, \textbf{b)} simulation output of stress components with non-trivial distribution from the spectral solver and corresponding network calculations from \textbf{c)} U-Net, and \textbf{d)} FNO}
    \label{fig:calculation 256}
\end{figure}
We trained and tested the ANNs at \(256\times256\) resolution with similar morphology (\(20\)-grain microstructures). Both ANNs successfully calculated the underlying morphology for all the stress components. The range of stresses for each component corresponded closely to the stress field from the spectral solver (see Figure \ref{fig:calculation 256}). We used normalized mean absolute error (NMAE) averaged over 100 unseen test cases to compare the average error in stress calculations from the networks. Computing NMAE(\%) for non-trivial stress components of \(P_{11},P_{22},P_{23},P_{32},P_{33}\) respectively, we found that FNO gave significantly lower values of \([0.282,0.287,0.389,0.388,0.266]\) than U-Net, which was \([2.15,1.49,1.41,1.42,1.55]\), i.e., an improvement by almost \(3.5-7.5\) times for different components. 

\begin{figure}[ht]
    \centering
    \includegraphics[height=5cm]{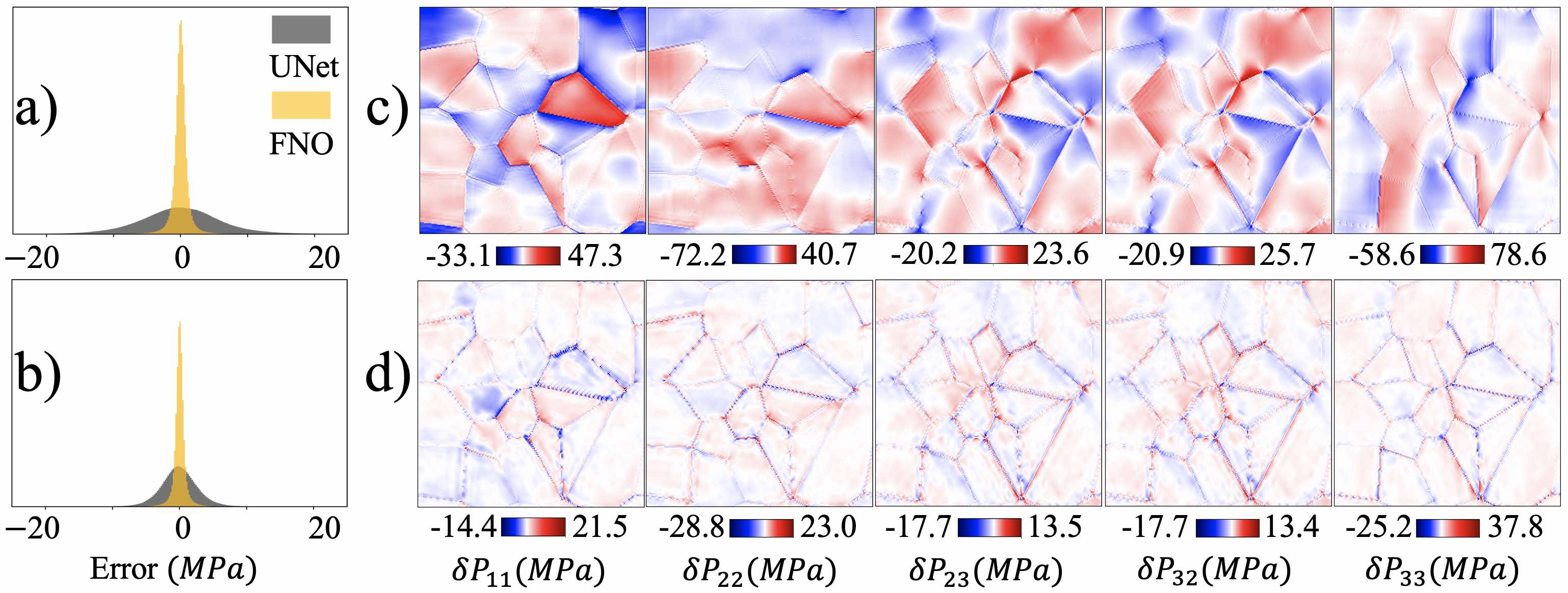}
    \caption{Error in stress field calculation for a test case with \(256\times256\) resolution and \(20\) grains \textemdash distribution of errors for \textbf{a)} normal components and \textbf{b)} shear components; calculation error in all stress components with non-trivial distribution for \textbf{c)} U-Net, and \textbf{d)} FNO.}
    \label{fig:Error 256}
\end{figure}

When we consider the point-wise errors in stress fields, for both ANNs there is a huge deviation from the error limits of typical fixed-point solvers. In fact, the errors are almost in the same range as the stress values themselves (see Figure \ref{fig:Error 256}). In the following, we discuss errors from different regions in the stress fields, i.e., errors from within the grains, and boundary errors, i.e., along the phase/grain boundaries and system boundary box. 

\paragraph{Bulk errors} We have broader error distribution in U-Net calculations with significant contributions from within the grains. Moreover, we observe blurred calculations for patches with higher stress gradients. The inability to capture these sharp features arises from the fact that the loss function allows for smoothening of sharp gradients as far as average errors are minimized. Additionally, the convolutional kernels have a tendency to learn intermediate mapping for grains and grain boundaries because these features are closely coupled in the spatial training data (grain boundaries are always surrounded by grain regions even for small square patches treated by kernels). This might lead to non-specialized learning for kernels, which can indiscriminately propagate errors over the entire domain. We can see this in Figure \ref{fig:Error 256}c, where U-Net calculations incur significant errors throughout the domain. On other hand, FNO learns linear maps in Fourier space. This corresponds to independently learning non-local features at different frequencies in real space. Some weights can specialize in high-frequency features like regions inside the grains, and some weights can specialize in low-frequency regions like grain boundaries. Hence, errors from within the grains are minimal, leading to a much narrower error distribution (see Figure \ref{fig:Error 256}b,d). The error is in fact limited to regions around the grain boundaries, indicating that very sharp gradients at the grain interfaces are still challenging to calculate for FNO.

\paragraph{Boundary errors} Both ANNs incur significant errors on the grain boundaries. In the case of FNO, this is the major source of errors. However, we do not observe any additional errors concentrated around the boundary box for either of the networks. Given that our data has periodic boundary conditions, boundary box errors will be less if the ANN is able to recognize periodicity. We incorporate this knowledge into U-Net by using periodic padding for convolutions. We found errors along the boundaries to emerge for U-Net if different padding, like zero padding, is used for the convolutions. However, this also limits the tANN's inference ability to microstructures with periodic boundary conditions. FNO inherently learns periodicity because of its parameterization in Fourier space, which is characterized by periodic kernels. 
\paragraph{Mechanical equilibrium violation} One of the key physical assumptions incorporated in micro-mechanics simulation is to maintain a mechanical equilibrium condition over displacement fields, which translates mathematically to divergence-free stresses, i.e., \(\nabla.\mathbf{P} = \mathbf{0}\). In either of the ANNs, we did not specify constraints to obey this assumption. Nonetheless, we found that U-Net incurs less error than FNO in terms of mechanical equilibrium violation. We computed the norm of averaged \(\nabla.\mathbf{P}\) deviation from zero vector per pixel over the domain and further averaged it over 10 cases. We found this norm to be \(1.02 \times 10^{-5}\), \(29.8 \times 10^{-5}\), and \(97.3\times10^{-5}\) for the spectral solver, U-Net, and FNO respectively, where the smaller value indicates a lower violation of the assumption. While these results are not comprehensive, they do indicate a trade-off between obeying mechanical equilibrium and reducing regression errors, considering that FNO gives lower regression errors with higher violation of mechanical equilibrium as compared to U-Net. Interestingly, the violation for FNO, even though larger than U-Net, was only limited to regions around grain boundaries, which wasn't the case for U-Net where violation within grains was also significant.

\paragraph{Speedups in comparison to the spectral solver}
The stress response in the spectral solver is computed for the applied load in multiple increments, and at every increment, it solves for the mechanical equilibrium of displacement fields using fixed-point iterations. On the other hand, FNO and U-Net calculate the stress response by performing computations in a single-pass feed-forward fashion through the trained network. Most of the computational burden is transferred to the training of the ANN, while the inference from a tANN happens at very little cost. We computed the speedups by testing the networks and running the simulation in spectral solver on a single core of 16-Core Intel Xeon W processor clocked at \(3.2\)GHz. Of course, using GPUs for testing ANNs would give even better scalability, but we did not use them so as to avoid the effects of parallelization overheads in numerical and ANN approaches. Compared to the numerical solver, we observed average speedups of \(\sim2500\) for FNO and \(\sim1000\) for U-Net. Here, we calculated/simulated stress fields with \(256\times256\) spatial resolution.

\subsection{Generalization with respect to spatial resolution} 
\label{subsec:super-res}

\begin{figure}[ht]
    \centering
    \includegraphics[height=6cm]{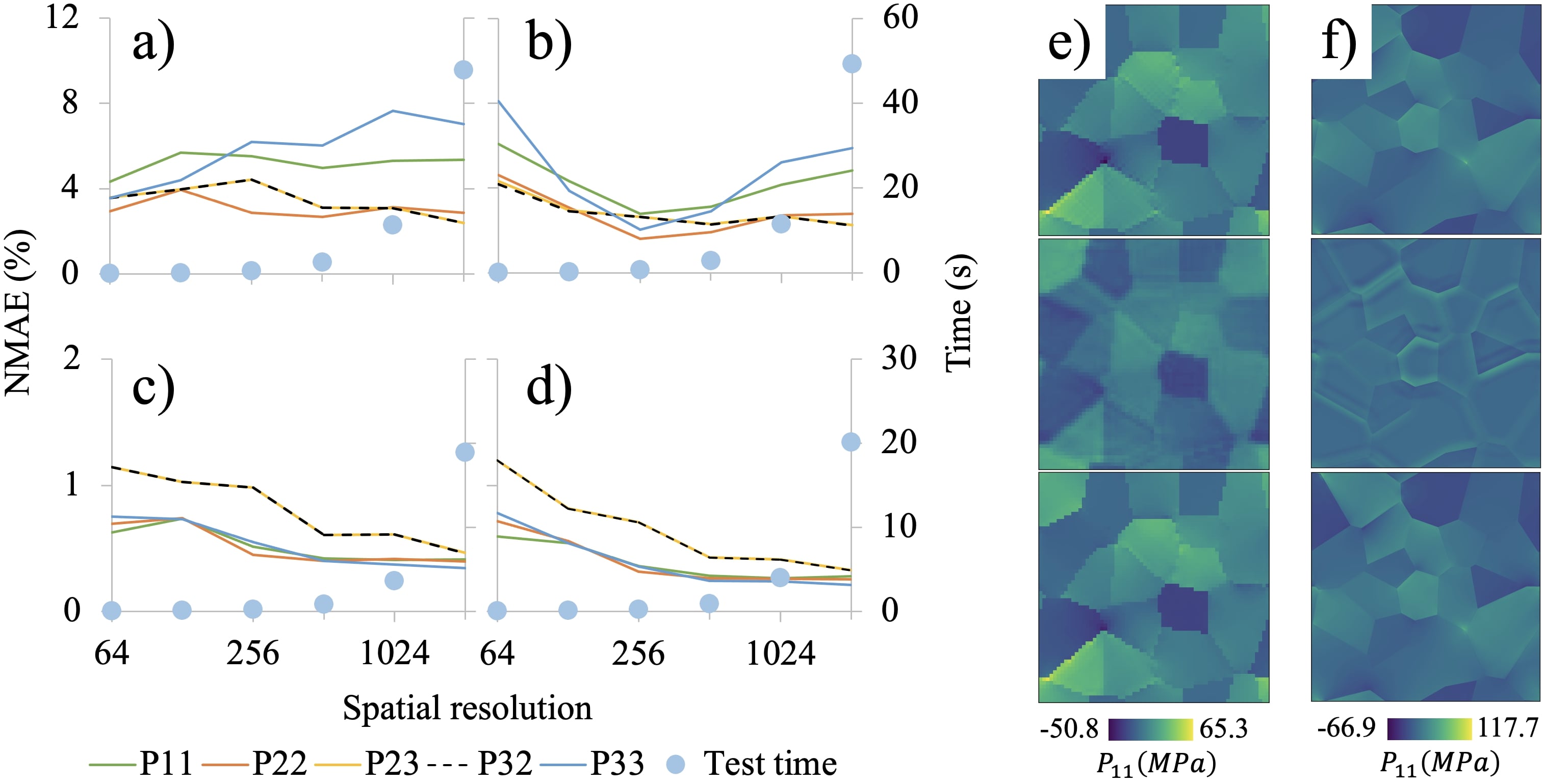}
    \caption{Normalized mean absolute errors (averaged over \(10\) cases) at different resolution after \textbf{a)} training U-Net with \(64\times64\) resolution and \textbf{b)} \(256\times256\) resolution, and \textbf{c)} training FNO with \(64\times64\) resolution and \textbf{d)} \(256\times256\) resolution. \(P_{11}\) stress field from the spectral solver, U-Net calculation, and FNO calculation (top to bottom) at \textbf{e)} \(64\times64\) and \textbf{f)} \(2048\times2048\) resolution.}
    \label{fig:Error vs Resolution}
\end{figure}

We tested U-Net and FNO at resolutions different from the training resolution, without varying the number of grains. In the case of U-Net, NMAE increased as the test resolution increased or decreased with respect to the training resolution (see Figure \ref{fig:Error vs Resolution}). The calculations at lower resolutions are blurred, whereas those at higher resolutions have ringing artifacts around the grain boundaries. This indicates that learning in U-Net is biased toward the resolution of the training dataset, as the errors are minimal only at the training resolution. Convolutional kernels learn relationships among neighboring pixels with translation invariance, but these relationships change as the system gets more/less resolved in terms of pixels, leading to a different data distribution altogether. For instance, the number of pixels between two-grain boundaries increases for a system if it gets more resolved. Convolutional kernels learn the pixel relationships at training resolution only and are incapable of generalizing them over other resolutions, hence leading to error-prone calculations.

Conversely, FNO calculations at higher or lower resolutions are sharper and free of artifacts. Further, NMAE increased for lower test resolutions and decreased for higher test resolutions (see Figure \ref{fig:Error vs Resolution}d). This monotonic trend can be reasoned by extending our observation about errors (see section \ref{subsec:stress field calculations} \textemdash it majorly comes from around the boundaries) to other resolutions. As the resolution is increased while keeping the grain size constant, a lower percentage of pixels represents the region around grain boundaries. Since these are the pixels that contribute most to the error, decreasing their percentage leads to lower mean errors. This reasoning also holds at a lower resolution, where a higher grain boundary percentage leads to higher average errors. Of course, a hidden assumption in this reasoning, which we find to be true, is that FNO learns the relationship between pixels independent of the system resolution. The assumption is also plausible because FNO learns frequencies of features across the domain, which is independent of spatial resolutions.  

Nevertheless, training FNO at a lower resolution is not as good as training it at a higher resolution. Errors for calculations at \(128\times128\) were \(1.25-1.35\) times higher for FNO trained at \(64\times64\) as compared to FNO trained at \(256\times256\) (see Figure \ref{fig:Error vs Resolution}c-d). Therefore, it is important for the system to be sufficiently resolved for FNO to optimally learn features at different frequencies.

\subsection{Generalization with respect to geometry}

One of the most desirable capabilities of an ANN would be its ability to calculate stress fields in different geometries than what it encountered during training, or more concisely, its generalizability to unseen geometries. To this end, we trained the networks on \(20\)-grain microstructures with \(256\times256\) resolution and tested them on three different types of morphology modification. Firstly, we varied the grain density at training resolution, which represents coarse- and fine-grained microstructures. We define the fineness and coarseness of the test case microstructure with respect to the microstructures in training data. Next, we drastically modified the morphology by adding a single inclusion of various shapes in the middle of the domain, without using grain-structure-like Voronoi tessellations. These cases are similar to the classic two-phase matrix-inclusion examples where Young's modulus of the two phases is very different, leading to soft (or hard) inclusion in a hard (or soft) matrix. However, as we were interested to study the effects of morphology alone on calculation accuracy, we did not introduce strong contrasts in Young's modulus of the two phases. Instead, we randomly selected it from a defined set as done for other material properties. Lastly, we changed the size of the boundary box in one dimension (fixing the other dimension) while scaling the number of grains accordingly to have similar grain shape and size, i.e., testing for varying aspect ratios of the boundary box.

\begin{figure}[t]
    \centering
    \includegraphics[height=7cm]{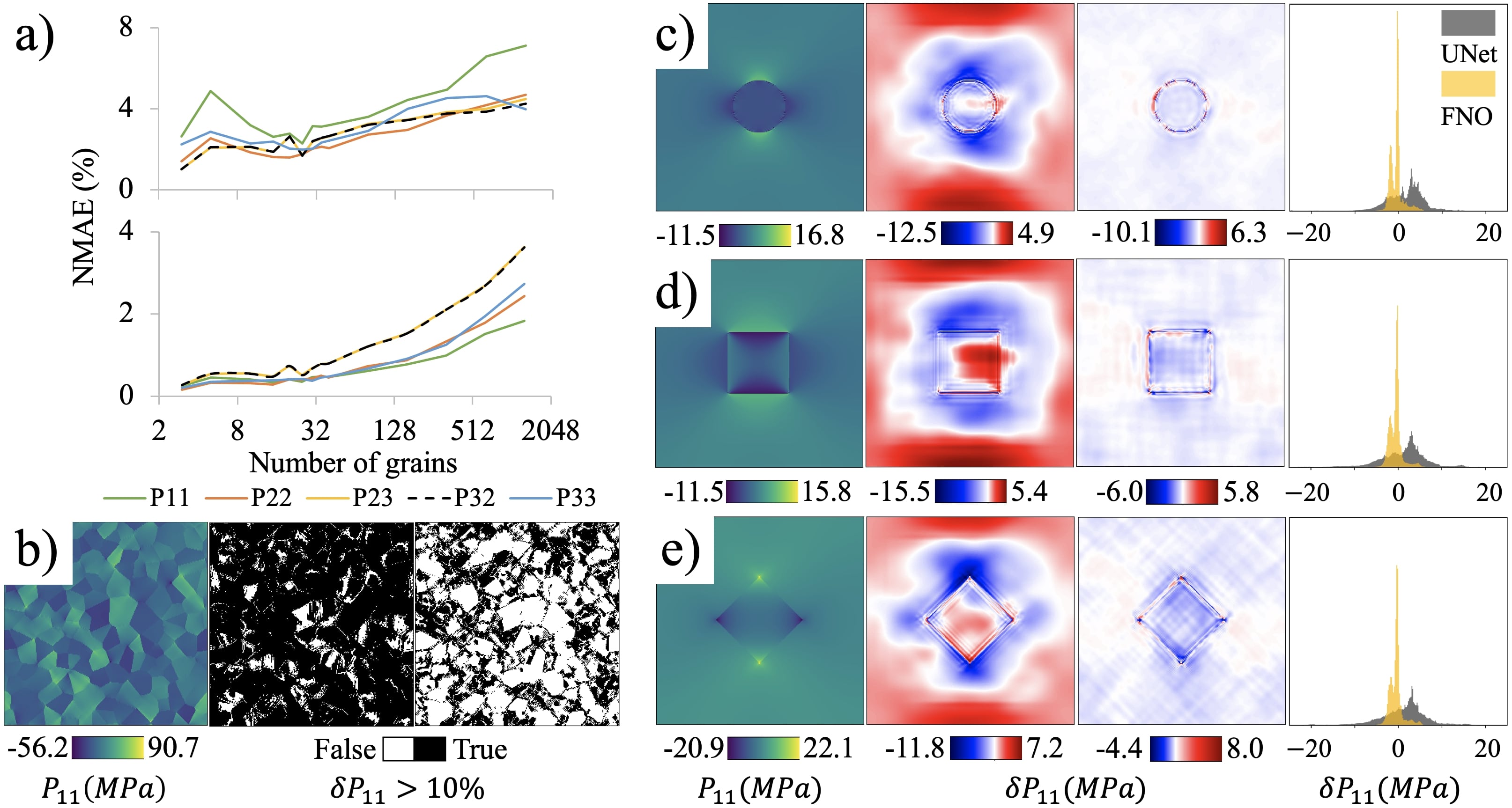}
    \caption{\textbf{a)} Normalized mean absolute errors (averaged over \(10\) cases) for calculations at \(256\times256\) resolution with varying grain sizes (plot for U-Net at the top and for FNO at the bottom). For \(P_{11}\) component, \textbf{b)} (\(\rightarrow\)) stress field from spectral solver for \(160\)-grain microstructure, distribution of errors \(>10\%\) for U-Net, and same for FNO. For \(P_{11}\) component, \textbf{c-e)} (\(\rightarrow\)) stress field from spectral solver for given inclusion-matrix microstructure, calculation error for U-Net, calculation error for FNO, and error distribution for both networks.}
    \label{fig:results_gs_ppt}
\end{figure}

\paragraph{Grain density} We increased (or decreased) the grain density by having more (or fewer) grains in the same-sized boundary box. Variation in grain density alters the spatial distribution of the grain boundaries. For accurate calculations in test cases with different grain densities, an ANN must be able to learn the correlation between material properties and stress response with less dependence on the spatial distribution of grain boundaries particular to the training set. We observed the learning in U-Net to be dependent on this distribution because the errors increased for both coarse-grained and fine-grained microstructures. That is, after optimizing the convolutional kernels for a given grain density, they struggle in inferring the stress fields for other grain densities. On other hand, errors in FNO were lower for coarse-grained and higher for fine-grained microstructures (see Figure \ref{fig:results_gs_ppt}a). Moreover, as observed before, most of the errors were concentrated around the grain boundaries and the errors within the grains are significantly lower (see Figure \ref{fig:results_gs_ppt}b). Therefore, errors in FNO depend on the frequency of prominent features like grain boundaries. While FNO is robust to subtle changes in this frequency as compared to that of the training dataset, drastic changes can quickly lead to high errors. 

\paragraph{Inclusions} With inclusion-matrix microstructure, we made the distribution of grain boundaries even farther from that of training data. Testing with circular, square, and diamond inclusions, we observed a wider error range for U-Net than FNO, with bulk errors dominating in U-Net calculations (see Figure \ref{fig:results_gs_ppt}c-e). This general trend also manifests in terms of NMAE for all stress components averaged over spatial fields of \(10\) cases (see Table \ref{tab:MAE ppt}), with U-Net performing specifically worse for \(P_{33}\) component. Interestingly, both ANNs fail to capture stress concentrations (especially at the vertices of diamond inclusion), which are the regions of great interest for material design decisions. 
\begin{table}[ht]
    \caption{Normalized Mean Absolute Error (in \%) averaged over \(10\) cases with randomly selected material properties of two phases for different shapes of inclusion.}
     \centering
     \begin{tabular}{p{1cm}p{2.5cm}p{0.8cm}p{0.8cm}p{0.8cm}p{0.8cm}p{0.8cm}}
       \toprule
       %\multicolumn{2}{c}{Part}                   \\
       %\cmidrule(r){1-2}
       & Inclusion shape & \(P_{11}\) & \(P_{22}\) & \(P_{23}\) & \(P_{32}\) & \(P_{33}\) \\
       \midrule
       U-Net & Circle  & 6.78 & 6.83 & 2.11 & 2.11 & 28.72  \\
       & Square  & 8.15 & 7.96 & 1.76 & 1.80 & 29.72 \\
       & Diamond  & 5.01 & 4.33 & 2.83 & 2.92 & 21.80 \\
       \midrule
       FNO & Circle  & 1.85 & 0.72 & 1.52 & 1.47 & 2.48  \\
       & Square  & 1.91 & 0.94 & 1.27 & 1.24 & 2.58  \\
       & Diamond  & 1.23 & 0.58 & 2.16 & 2.11 & 1.88 \\
       \bottomrule
     \end{tabular}
     \label{tab:MAE ppt}
\end{table}
\paragraph{Boundary box aspect ratio}
So far, we trained and tested the networks for square boundary boxes with an aspect ratio of 1. However, in practical applications, there may arise situations where the boundary box is rectangular. It will be an added advantage if an ANN trained on square boundary cases can successfully calculate stress response in geometries with rectangular boundaries. To this end, we trained the ANNs with \(256\times256\) resolution 20-grain microstructures and tested them for two different variations of the boundary box. We fixed one dimension at \(256\) pixels and increased/decreased the number of pixels in the other dimension resulting in aspect ratios greater/smaller than \(1\) with respect to the square training resolution. The number of grains in varying dimension was altered such that the grain size and morphology is similar to the training dataset. Only the system becomes bigger/smaller in one dimension.
 
\begin{figure}[ht]
    \centering
    \includegraphics[height=8cm]{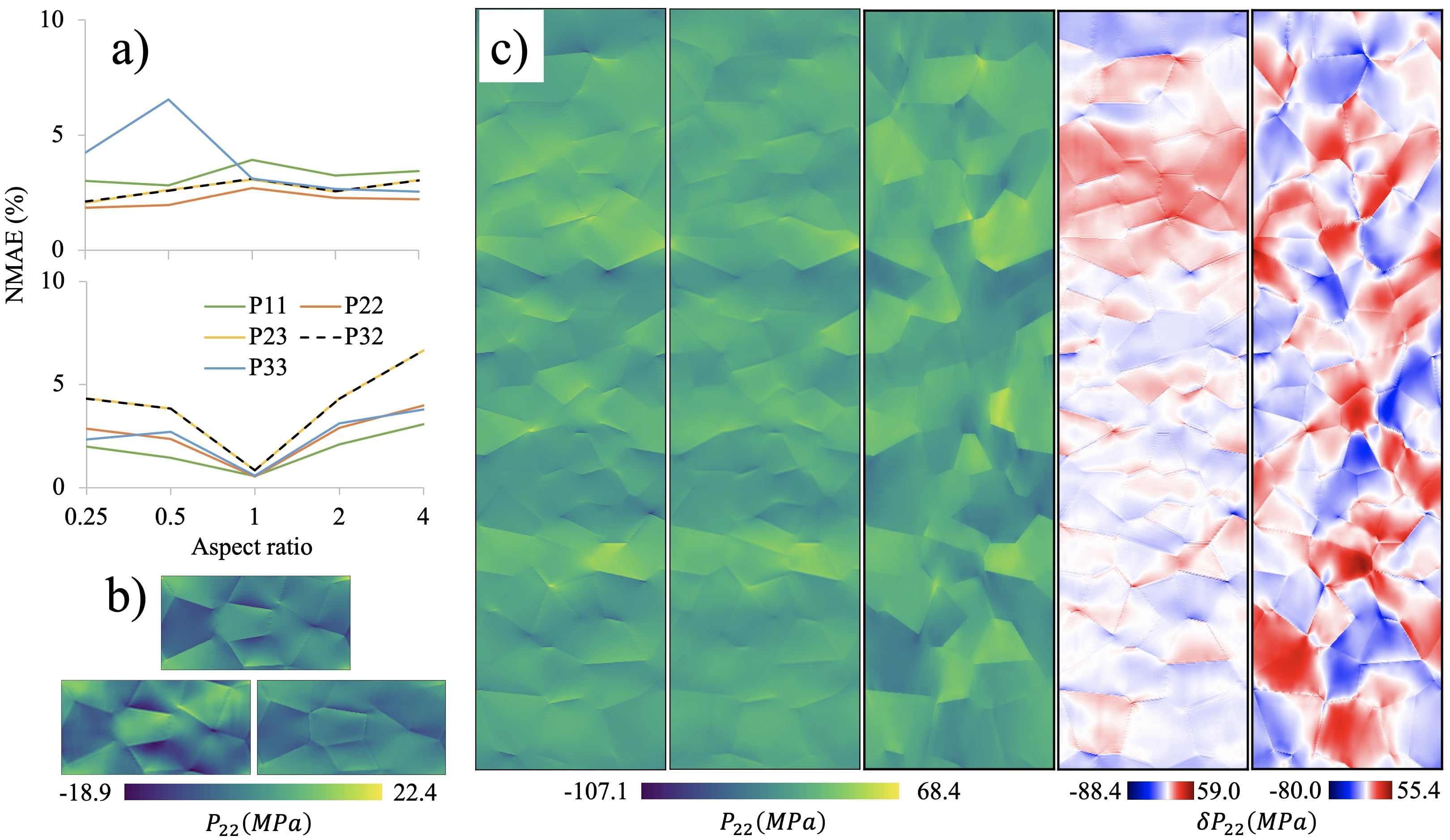}
    \caption{\textbf{a)} Error in \(P_{22}\) field calculation (averaged over \(5\) cases) for different aspect ratios of boundary box, where one dimension is fixed at \(256\) pixels and the other dimension is varied from \(64\) to \(1024\) pixels (plot for U-Net at the top and for FNO at the bottom). Here, the average grain size is similar to training. \textbf{b)} (anticlockwise) Stress field from the spectral solver, U-Net calculation, FNO calculation for boundary box with aspect ratio \(256:128\). \textbf{c)} (\(\rightarrow\)) Stress field from the spectral solver, U-Net, FNO, and corresponding error fields from U-Net and FNO. The boundary box in this case has an aspect ratio of \(256:1024\).}
    \label{fig:results_aspect_ratio}
\end{figure}

In the case of U-Net, the errors remain almost the same for most of the stress components (see Figure \ref{fig:results_aspect_ratio}). This is reasonable as the convolutional kernel operations are localized over small square patches, therefore the change in the aspect ratio of the boundary box does not alter their behavior. However, in the case of FNO, the Fourier layers compute feature maps in a non-local fashion based on the frequency of features. As one of the dimensions grows/shrinks, it is accompanied by a change in the frequency of features as seen by Fourier modes. For example, when the dimension grows from \(256\) to \(1024\) pixels, one period of the first Fourier mode contains \(4\) times more grains than before. This change in the frequency of features makes the underlying distribution different than the training dataset, leading to higher inference errors. Hence, the averaged errors for FNO increased for both the variations of the boundary box.

\section{Summary and conclusions}
\label{sec:ConSum}

In this work, we implemented U-Net (CNN-based approach) and FNO (operator-learning approach) as surrogate models to calculate stress response in synthetic polycrystalline microstructures. We trained these ANNs to generate fields under tensile loading for non-trivial stress components and found both the ANNs to give tremendous speedups as compared to the spectral solver. We systematically compared the errors in stress fields generated by FNO and U-Net after training them with 20-grain microstructures having a square domain. These comparisons were based on accumulated average errors (NMAE) and local errors for different test cases. One of the key differences between the two approaches is that local errors in FNO are limited to regions around grain boundaries, whereas the errors in U-Net come from within the grains as well. As a consequence, the NMAE is \(3.5-7.5\) times lower in the former than the latter for different components accompanied by narrower distribution around zero. 

FNO is able to generalize better with variations in spatial resolution. The average errors scaled with the percentage of the grain boundary region, i.e., they increased at a lower resolution and decreased at a higher resolution. Whereas the errors in U-Net only increased with both types of variation in the spatial resolution. We also observed ringing artifacts in U-Net at higher resolutions. FNO is robust to small variations in grain density of the microstructure, however, larger variations lead to a steep increase in errors. Conversely for U-Net, even small variations in grain density lead to higher errors. For two-phase matrix-precipitate microstructures, the error distribution for FNO is narrower than U-Net's, however, both the ANNs failed to capture stress concentrations around sharp corners. On other hand, U-Net is more robust to variation in the aspect ratio of the boundary box than FNO. The errors in FNO are dependent on the frequencies of the prominent features, and as the grain density increases or the aspect ratio is made farther than \(1\) while maintaining the grain density, these frequencies are altered. As a consequence, in such cases, FNO incurs high inference errors. 

\begin{figure}[ht]
    \centering
    \includegraphics[height=6cm]{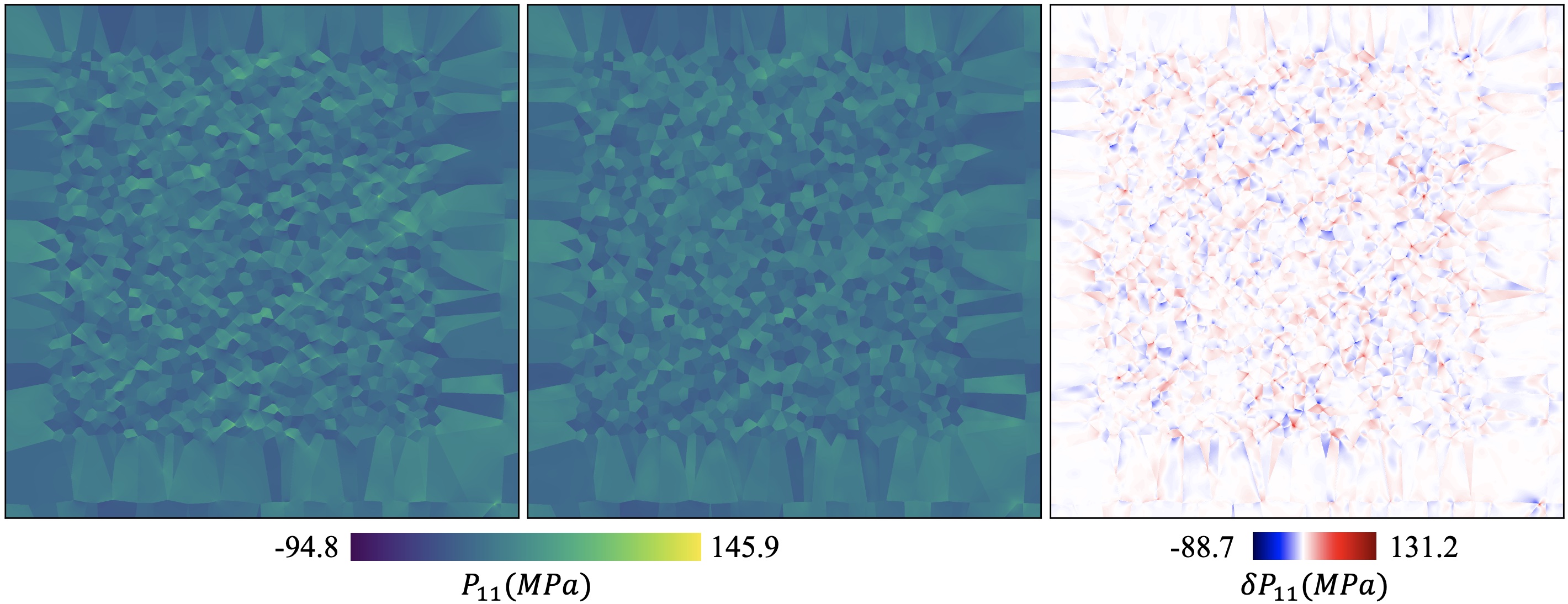}
    \caption{Testing FNO with \(1280\)-grain microstructure (grains having different shape and sizes) at \(2048\times2048\) resolution, which was trained with \(20\)-grain microstructures (all grains having similar sizes) at \(256\times256\) resolution. (\(\rightarrow\)) Stress field \(P_{11}\) from the spectral solver, FNO, and error in FNO calculation. The mean absolute error is \(2.78\)MPa with a standard deviation of \(3.41\)MPa.}
    \label{fig:FNO calculation geom and resolution}
\end{figure}

Through this work, we show the key differences between the two popular ANN approaches and highlight their pros and cons. FNO's robustness to variation in spatial resolution is highly favorable as it allows to faithfully generate higher resolution simulations after training at lower resolutions. As an example, we generated a stress field at 8-times the resolution and 640-times the number of grains as compared to the training dataset (see Figure \ref{fig:FNO calculation geom and resolution}), which is qualitatively comparable to the output of the spectral solver. On other hand, U-Net's robustness to variation in the aspect ratio of the boundary box allows for more freedom in the shape of the domain. An ideal ANN would combine these two properties. Further developments of ANNs should also focus on retaining the sharpness of strong stress gradients, especially at the regions of stress concentration and grain boundaries. Additionally, as discussed in Section \ref{subsec:stress field calculations}, both of the ANNs are prone to violation of physical constraints like mechanical equilibrium. Building these constraints within the ANN architecture or the loss function can mitigate this issue. A fruitful endeavor could also be to combine ANN approaches with numerical approaches like spectral solvers, where calculations from ANNs can be used as an initial condition for solvers. This can help reduce the iterative load of numerical solvers while retaining the physical correctness of the simulations. 

%\section*{Acknowledgments}
%This was supported in part by......

% \printglossary[]
\bibliographystyle{unsrt}  
\bibliography{references}  

\end{document}